\documentclass[a4paper,12pt]{article}
\usepackage{epsfig}
\usepackage{psfrag}
\usepackage{graphicx}
\usepackage{cite}
\usepackage{a4wide}

\begin{document}
\vskip 1.8cm
\centerline{\Large \bf Black Holes and the Holographic Principle } 
\vskip 1.4cm \centerline{\bf L\'arus Thorlacius} \vskip .4cm
\centerline{\sl Science Institute, Dunhaga 3, 107 Reykjavik, Iceland}
\centerline{\tt lth@hi.is} 
\begin{abstract}
\noindent
This lecture reviews the black hole information paradox and briefly
appraises some proposed resolutions in view of developments in
string theory. It goes on to give an elementary introduction to the 
holographic principle.
\end{abstract}
\setcounter{footnote}{0} \setcounter{page}{1}

\section{Introduction}

The theory of black holes involves a subtle interplay between gravity and 
quantum physics. Semiclassical arguments indicate that the time evolution 
of a system, where a black hole forms and then evaporates, cannot be 
governed by the standard postulates of quantum mechanics.  If the black 
hole forms by gravitational collapse from an initial matter configuration 
that is nonsingular and described by a pure quantum state, and if Hawking 
radiation is truly thermal, then the formation and evaporation process 
evolves a pure state into a mixed one, in violation of quantum mechanical 
unitarity. Alternatively, unitarity may be maintained in black hole 
evolution but at the price of giving up locality at a fundamental level.

The long-standing debate regarding these issues, initiated by 
Hawking \cite{Hawking:1975sw,Hawking:1976ra}, 
played a key role in the development of
the {\it holographic principle} \cite{'tHooft:1993gx,Susskind:1995vu}. 
This radical principle 
goes beyond black hole physics. It concerns the number of degrees of freedom 
in nature and states that the entropy of matter systems is drastically 
reduced compared to conventional quantum field theory. This claim is
supported by the covariant entropy bound \cite{Bousso:1999cb} 
which is valid in
a rather general class of spacetime geometries. The notion of holography 
is well developed in certain models and backgrounds, in particular in the
context of the adS/cft correspondence. A more general formulation is 
lacking, however, and the ultimate role of the holographic principle in 
fundamental physics remains to be identified.

\section{Black hole evolution}

\begin{figure}
\begin{center}
\psfrag{rnull}{$r=0$}
\psfrag{P}{$P$}
\psfrag{scriplus}{ }
\psfrag{scriminus}{ }
\psfrag{i0}{$r=\infty$}
\includegraphics[width=5cm]{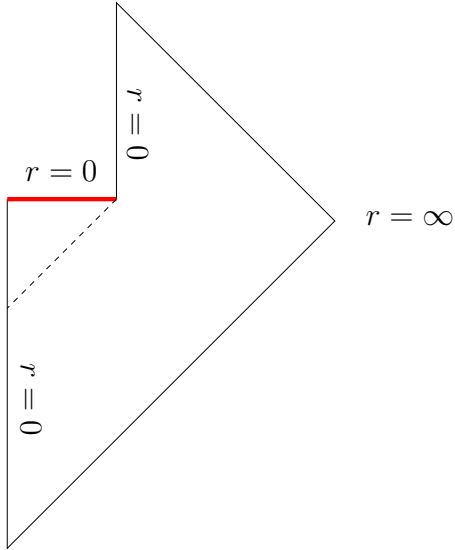}
\end{center}
\caption{\footnotesize
Penrose diagram for a semiclassical black hole geometry. At early and 
late times the geometry approaches that of Minkowski spacetime.}
\label{figone}
\end{figure}

Let us start by reviewing the basic ingredients that go into the 
black hole information paradox.  We consider a black hole formed in 
gravitational collapse and let us assume that the initial matter 
configuration is approximately spherical and sufficiently diffuse so that 
spacetime curvature is everywhere small at early times.  The subsequent 
evolution, including the formation and evaporation of the black hole, is
then well represented by the Penrose diagram in Figure~\ref{figone}. 
The diagram assumes exact spherical symmetry. Only radial and time 
coordinates are displayed, with each point in the diagram 
representing a transverse two-sphere, whose area depends on the
radial coordinate. 

Penrose diagrams are a useful means of portraying the global 
geometry and causal properties of a given spacetime. Without 
going into details we note two key properties. 
First of all, the coordinates are chosen in such a way that radially 
directed light-rays correspond to straight lines oriented at $\pm 45$
degrees to the vertical axis. The second feature is that, following
a conformal mapping that brings infinity to a finite point, the entire 
spacetime geometry is represented by a finite region.  As a result, 
the causal relationship between any two events is easily read off a 
Penrose diagram but proper distances in spacetime are not faithfully 
depicted. 

The event horizon, shown as a dotted line in Figure~\ref{figone},
defines the boundary of the black hole region, inside which timelike
observers cannot avoid running into the future singularity.
An important feature of black holes is that local gravitational 
effects are extremely weak at the event horizon of a large black
hole. In fact, any curvature invariant formed from the Riemann
tensor will go as an inverse power of the black hole mass.
For a Schwarzschild black hole, for example, one finds
\begin{equation}
R_{\mu\nu\lambda\sigma}R^{\mu\nu\lambda\sigma} 
= {3\over 4 l^2_{\rm Pl}}\, {M^4_{\rm Pl}\over M^4}
\end{equation}
at the event horizon. This observation forms the basis of the
semiclassical approach to black holes, which assumes
that only low-energy physics is involved in the formation and
evaporation of a black hole, except in the region near
the singularity, and that away from this region physics can be described 
by a local effective field theory. The detailed construction of such an 
effective field theory for black hole evolution presents a formidable 
and unsolved technical problem, but let us for the moment assume that 
such a theory can be found and sketch the argument for information loss 
in black hole evolution.

\subsection{Semiclassical information loss}

\begin{figure}
\begin{center}
\psfrag{sin}{$\Sigma_{\textrm{\footnotesize in}}$}
\psfrag{sout}{$\Sigma_{\textrm{\footnotesize out}}$}
\psfrag{sext}{$\Sigma_{\textrm{\footnotesize ext}}$}
\psfrag{sbh}{$\Sigma_{\textrm{\footnotesize bh}}$}
\psfrag{rnull}{$r=0$}
\psfrag{P}{\footnotesize P}
\includegraphics[width=5cm]{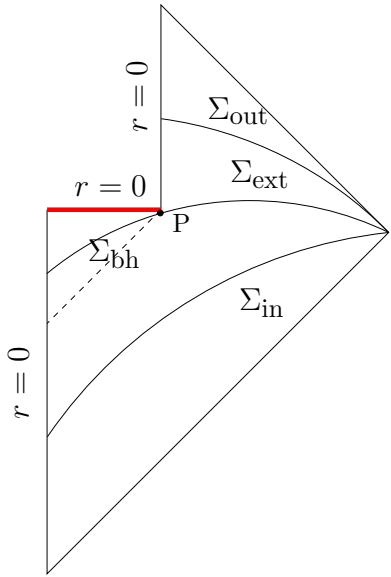}
\end{center}
\caption{\footnotesize
Cauchy surfaces used in the argument for information loss in
black hole evolution.}
\label{figtwo}
\end{figure}

The first step is to choose appropriate spatial slices through the 
black hole spacetime to provide a set of Cauchy surfaces for the
quantum evolution of our system. The initial slice is taken to lie in 
the asymptotic past where spacetime is approximately flat and contains 
a diffuse distribution of matter that will later undergo gravitational 
collapse. The final slice, at asymptotically late times, contains a 
long train of outgoing Hawking radiation and possibly also a Planck
mass remnant of the black hole.\footnote{The final stage of the evaporation 
process is governed by Planck scale physics, which we have limited 
knowlegde of, and thus we cannot preclude the existence of black hole
remnants. The important question for the information problem is not 
whether remnants exist but rather how many distinct remnant states are 
possible.  We discuss black hole remnants in Section~\ref{remnantsec}.} 
It turns out to be possible to choose spatial slices at an intermediate 
stage in such a way that they lie partly inside the black hole region 
and partly outside, as indicated in Figure~\ref{figtwo}.  
For a large enough black hole this can be done in such a way that the 
following two requirements are met:
\begin{enumerate}
\item
There are Cauchy surfaces that intersect the worldlines of the infalling 
matter inside the black hole but also the worldlines of most of the 
outgoing Hawking radiation that is emitted during the black hole lifetime.
\item
These spatial slices avoid the region of strong curvature near the
black hole singularity and are also smooth in the sense that their
extrinsic curvature is everywhere small.
\end{enumerate}
An explicit construction of a family of {\it nice slices\/} of this type
is for example given in \cite{Lowe:1995ac}.

The semiclassical theory of black hole evolution rests on the assumption
that, given such a family of Cauchy surfaces, the dynamics of the
combined matter and gravity system is governed by a low-energy
effective field theory and that no Planck scale effects enter into
the physics except near the curvature singularity. The argument for
information loss is based on the existence of this effective field theory
but not on its detailed form. 

Suppose an initial configuration of collapsing matter in a weakly curved 
background is described by a pure quantum state 
$\vert\psi(\Sigma_{\textrm{\footnotesize in}})\rangle$ 
defined on the surface $\Sigma_{\textrm{\footnotesize in}}$ in
Figure~\ref{figtwo}. The Hamiltonian of the effective field theory 
generates a linear evolution of this state into another pure state 
$\vert\psi(\Sigma_{\textrm{\footnotesize P}})\rangle$ on 
the surface $\Sigma_{\textrm{\footnotesize P}}$,
which is partially inside and partially outside the black hole region.
The inside and outside portions of $\Sigma_{\textrm{\footnotesize P}}$, 
denoted by $\Sigma_{\textrm{\footnotesize bh}}$ 
and $\Sigma_{\textrm{\footnotesize ext}}$ respectively, are spacelike
separated and as a result all observables in the effective field
theory that have support on $\Sigma_{\textrm{\footnotesize ext}}$
commute with observables that have support on 
$\Sigma_{\textrm{\footnotesize bh}}$. The state on 
$\Sigma_{\textrm{\footnotesize P}}$ is therefore an element of a tensor
product Hilbert space,
\begin{equation}
\vert\psi(\Sigma_{\textrm{\footnotesize P}})\rangle \in
\mathcal{H}_{\textrm{\footnotesize bh}} \otimes 
\mathcal{H}_{\textrm{\footnotesize ext}} \, .
\end{equation}
Observers outside the black hole have no access to the part of 
$\vert\psi(\Sigma_{\textrm{\footnotesize P}})\rangle$ that is inside
the black hole and ultimately runs into the singularity. As a result 
we are instructed to trace over states in 
$\mathcal{H}_{\textrm{\footnotesize bh}}$
giving rise to a mixed state density matrix on 
$\mathcal{H}_{\textrm{\footnotesize ext}}$. This mixed state will 
then evolve into another mixed state on the late time Cauchy surface
$\Sigma_{\textrm{\footnotesize out}}$.

We now have a paradox on our hands, for if the entire process of black
hole formation and evaporation is to preserve unitarity, then the final
configuration of the system must be described by a pure quantum state
$\vert\psi(\Sigma_{\textrm{\footnotesize out}})\rangle$. This final
state is obtained from the initial state by a unitary S-matrix,
\begin{equation}
\vert\psi(\Sigma_{\textrm{\footnotesize out}})\rangle 
= S \,
\vert\psi(\Sigma_{\textrm{\footnotesize in}})\rangle \,,
\end{equation}
where $S\,S^{\dagger}= {\bf 1}$.
This relation can in principle be inverted to express the initial 
state in terms of the final one,
\begin{equation}
\vert\psi(\Sigma_{\textrm{\footnotesize in}})\rangle
= S^\dagger \,
\vert\psi(\Sigma_{\textrm{\footnotesize out}})\rangle \,,
\end{equation}
In other words, in a unitary theory the final quantum state carries all 
information that is contained in the initial state. The semiclassical
argument, on the other hand, resulted in a mixed final state from which
there is no way to recover the initial state. In other words, quantum 
information is lost in semiclassical black hole evolution.

\subsection{Proposed resolutions}

There are a number of ways to respond to this paradox. Let us briefly
review the three main proposals. 

\subsubsection{Information loss}

Hawking advocated that the above 
semiclassical argument should be taken at face value and that an added 
fundamental uncertainty is to be incorporated into quantum physics 
when gravitational effects are taken into account 
\cite{Hawking:1976ra}.
He also had a concrete proposal involving a modified set of axioms 
for quantum field theory that allows pure states to evolve into mixed
ones. In Hawking's formalism the unitary S-matrix of conventional quantum
field theory, which maps an inital quantum state to a final quantum 
state, is replaced by a superscattering operator \$, which maps an 
initial density matrix to a final density matrix, and processes involving
black hole formation, or even virtual processes involving gravitational
fluctuations, give rise to superscattering that mixes quantum states. 

This proposal was criticised by a number of authors
\cite{Banks:1984by,Ellis:1984jz}. 
In particular, Banks {\it et al.} argued
that Hawking's density matrix formalism is equivalent to conventional 
quantum field theory coupled to randomly fluctuating sources everywhere
in spacetime. As a result the theory does not have empty vacuum as it's
ground state but rather a thermal configuration at the Planck 
temperature \cite{Banks:1984by}. This is clearly phenomenologically 
unacceptable 
but it should be noted that the argument rests on certain technical 
assumptions and conceivably a loophole may be found to avoid the thermal 
disaster. The early critisicm appears to have put a stop to further 
developments in this direction and the general view is that Hawking's 
density matrix formalism is not a viable option for resolving the 
information paradox. This, of course, does not rule out a theory 
incorporating information loss being developed in the future. 

At the moment, however, our best candidate for a theory of quantum gravity 
is superstring theory and this theory does not favor information loss.
In it's original formulation, string theory is an S-matrix theory and
as such it is manifestly unitary. On the other hand, the original 
formulation of string theory really only amounts to a perturbative 
prescription for scattering amplitudes and is not adequate for describing 
macroscopic processes such as the formation of a large mass black hole.
On the other hand, we now have non-perturbative formulations of string
theory both in an asymptotically flat background \cite{Banks:1997vh} and in
anti-de Sitter spacetime \cite{Maldacena:1998re}, where the 
gravitational dynamics has a dual description which is unitary. 
Admittedly, in both cases the duality is founded on conjectures 
that are unproven, and will be difficult to prove in all generality because 
they involve weak-strong coupling dualities, but supporting evidence has 
been pouring in for several years now, building a strong case.
We will return to this in section \ref{adscft}.

\subsubsection{Black hole remnants}
\label{remnantsec}

An alternative viewpoint, put forward by Aharonov {\it et al.} 
\cite{Aharonov:1987tp}, 
is that black holes do not completely evaporate but rather leave behind  
remnants that are stable or extremely long-lived.  Quantum mechanical 
unitarity is then maintained by having the black hole remnant carry 
information about the initial quantum state of infalling matter that 
forms the black hole.

If we assume that Hawking's semiclassical calculation of particle emission
remains valid until the remaining black hole mass approaches the Planck
scale and that none of the initial information goes out with the Hawking
radiation then the mass of a black hole remnant can be no more than a few 
times the Planck mass and there needs to be a distinct remnant for every
possible initial state. As a result, the density of these remnant states 
at the Planck energy must be virtually infinite and this leads to severe 
phenomenological problems if the remnants behave at all like localized
objects. Their effect on low-energy physics could then be described in
terms of an effective field theory and contributions from virtual remnant 
states would dominate almost any quantum process one might consider. 
Even if the amplitude for producing any given Planck mass remnant as an 
intermediate in, say, $e^+e^-$ scattering at a collider is extremely small, 
the infinite density of such states would nevertheless make remnants the 
dominant channel. An infinite density of states also leads to a divergent 
pair production rate of remnants in weak background fields and to 
divergent thermal sums. Since these effects are not observed
either the information carried by a black hole is not left behind in a 
Planck scale remnant or such remnants are described by very unconventional 
laws of physics at low energy. 

About ten years ago some remnant models were suggested, where these 
pathologies were to be avoided by accommdating the high density of states 
in a large internal volume carried by the remnant and connected to the rest 
of spacetime via a Planck scale throat region 
\cite{Banks:1993mi}. Although the
models had some success and served as a warning against drawing too firm
a conclusion from arguments based on effective field theory, they have
not been developed further. A major reason for this can be traced to 
subsequent developments in string theory. With the advent of string duality
and branes the basic degrees of freedom of string theory have more or less
been identified, and they do not include exotic black hole remnants
at the Planck scale. 

One of the triumphs of string theory in the nineties was the microphysical 
explanation and direct calculation of the Bekenstein-Hawking entropy of 
certain extremal and near-extremal black holes \cite{Strominger:1996sh}.  
The entropy is obtained by counting configurations of strings and branes 
that carry the same charges as the black hole in question. Such counting 
is only reliable for a weakly coupled collection of strings and branes 
in flat spacetime which bears little resemblance to the strongly curved 
geometry of a black hole, but for a configuration that corresponds to an 
extremal black hole one appeals to extended supersymmetry and 
non-renormalization theorems to argue that the counting will, in fact, 
also hold at strong coupling where the system is more appropriately 
described as a black hole.  The notion that, by moving around inside the 
parameter space of the theory, one can find a dual description of black
holes in terms of weakly coupled (highly excited) strings and 
branes \cite{Horowitz:1997nw} leaves no room for a divergent density of 
remnant states at the Planck energy.  

\subsubsection{Black hole complementarity}

A third possibility, pioneered by Page \cite{Page:1980qm} and 
't~Hooft \cite{tHooft:1990fr}, 
is that Hawking radiation is not exactly thermal but in fact carries all 
the information about the initial state of the infalling matter.  This 
information must then be encoded in subtle correlations between quanta 
emitted at different times during the evaporation process because even
if the formation and evaporation process as a whole is governed by a 
unitary S-matrix the radiation emitted at any given moment will appear
thermal.  Detecting the information would require statistical analysis 
of a large number of observations made on an ensemble of black holes 
formed from identically prepared initial states. This is a conservative
viewpoint in that it assumes unitarity in all quantum processes, also
when gravitational effects are taken into account, but it leads to a 
novel view of spacetime physics and requires us to give up the notion
of locality at a fundamental level.

The question is whether the infalling matter will give up all information
about its quantum state to the outgoing Hawking radiation or whether the
information gets carried into the black hole. If the information is 
imprinted on the Hawking radiation then it must also be removed from
the infalling matter for otherwise we would have a duplication of the
information in a quantum state in violation of the linearity of quantum
evolution. 
We can compare this to the more conventional information loss when a 
book is burned. All the information that was originally contained in
the book
can in principle be learned from measurements on the outgoing smoke
and radiation, but at the same time this information is no longer to be 
found in the remains of the book. In this case, however, it is a well
understood microphysical process that removes the information from
the book and transfers it to the outgoing radiation, whereas matter in
free fall entering a large black hole does not encounter any disaster
before passing through the event horizon.

The principle of {\it black hole complementarity} \cite{Susskind:1993if} 
states that 
there is no contradiction between outside observers finding information
encoded in Hawking radiation, and having observers in free fall pass
through the event horizon unharmed. The validity of this principle 
requires matter to have unusual kinematic properties at very high 
energy but it does not conflict with known low-energy physics. 
Contradictions only arise when we attempt to directly compare the 
physical description in widely different reference frames. The laws of
nature are the same in each frame and low-energy observers in any
single frame cannot establish duplication of information 
\cite{Susskind:1994mu}.

\subsubsection{The stretched horizon}

In order to illustrate the concept of black hole complementarity it is 
useful to have a physical picture of the evaporation process 
in the outside frame.  For some time astrophysicists have made use of
the membrane paradigm of black holes to describe the classical physics 
of a quasistationary black hole \cite{Thorne:1986iy}. From the point of 
view of outside observers the black hole is then replaced by a 
{\it stretched horizon}, which is a membrane placed near the event
horizon and endowed with certain mechanical, electrical and thermal
properties. This description is dissipative and irreversible in time.
One does not have to be specific about the location of the stretched
horizon as long as it is close to the event horizon compared to the
typical length scale of the astrophysical problem under study.

In the context of black hole evaporation one goes a step further 
and views the classical stretched horizon as a coarse grained 
thermodynamic description of an underlying microphysical system, a 
quantum stretched horizon, located a Planck distance outside the event 
horizon and with a number of states given by $\exp{(A/4)}$, where $A$ 
is the black hole area in Planck units \cite{Susskind:1993if}.
The nature of the microphysics involved was left unspecified 
in \cite{Susskind:1993if} but it
was later suggested that the dynamics of the stretched horizon might 
be explained in the context of string theory 
\cite{Susskind:1993ki}. The sticky
part is that, in order to implement black hole complementarity, we
have to stipulate that this membrane, or stretched horizon is 
absent in the reference frame of an observer entering the black hole
in free fall.

The evaporation of a large black hole is a slow process and, for the
purpose of our discussion here, the evolving geometry may be approximated
by a static Schwarzschild solution, 
\begin{equation}
ds^2 = -(1-{2M\over r})dt^2 +(1-{2M\over r})^{-1}dr^2 +r^2d\Omega^2 \,.
\end{equation}
Provided the mass $M$ is sufficiently large, an observer in free
fall will not experience any discomfort upon crossing the event horizon
at $r=2M$ but a so-called fiducial observer, who is at rest with 
respect to the Schwarzschild coordinate frame, will be bathed in 
thermal radiation at a temperature that depends on the radial position,
\begin{equation}
T(r) = {1\over 8\pi M}(1-{2M\over r})^{-1/2} \,.
\label{temperature}
\end{equation}
This temperature diverges near the black hole, 
$T\approx (2\pi \delta)^{-1}$, where $\delta$ is the proper distance 
between the fiducial observer and the event horizon.
The radiation can be attributed to the acceleration required to remain
stationary at fixed $r$, which diverges as $\delta\rightarrow 0$.
In the spirit of the membrane paradigm the radiation can also be viewed as 
thermal radiation emanating from a hot stretched horizon. The region 
nearest the event horizon, where the temperature (\ref{temperature}) 
formally diverges, is then replaced by a Planck temperature membrane 
located a proper distance of one Planck length outside the event horizon.

The stretched horizon is then the source of Hawking radiation as far as
outside observers are concerned. It emits Planck energy particles but
most of these will fall back into the black hole. Those who escape its
gravitational pull predominantly carry low angular momentum and are
redshifted to energies of order the Hawking temperature by the time they 
reach the asymptotic region. In this view, Hawking radiation is the
enormously redshifted glow from a Planck scale inferno at the stretched 
horizon. It carries all information about the initial state of the
matter that formed the black hole, albeit in a severely scrambled form.

No infalling observer survives the encounter with the hot membrane. 
In fact nothing ever enters the black hole in the reference frame of 
outside observers, including the matter that formed the black hole in 
the first place. It is familiar from classical general relativity that
matter falling into a black hole appears to slow down as it approaches
the horizon and fades out of view due to the gravitational redshift.
This is after all where the term black hole comes from. In the quantum
membrane picture, the infalling matter runs into the stretched horizon, 
gets absorbed into it and thermalized, and is then slowly returned back 
out along with the rest of the black hole as it evaporates.

This scenario conflicts with the notion that infalling observers feel
no ill effects upon passing through the event horizon of a large
black hole. The analysis of several gedanken experiments, designed 
to expose possible contradictions between the experience of infalling
observers and the description in the outside reference frame, led to
the conclusion that the apparent contradictions could in each case
be traced to assumptions about physics at or above the Planck 
scale \cite{Susskind:1994mu}. This observation does not by itself resolve the 
information problem but it challenges the underlying assumption of the 
semiclassical approach that the paradox can be posed in terms of 
low-energy physics alone without any reference to the Planck scale.

If we take the principle of black hole complementarity at face value
we have to accept a radically new view of spacetime physics. The notion
of a local event, that is invariant under coordinate transformations, 
is central to general relativity. According to black hole complimentarity
observers in different reference frames can totally disagree about the 
location of the rather significant event where an observer falling into 
a black hole meets his or her end. The proper distance between the 
event horizon and the final singularity is proportional to the 
black hole mass and can therefore be arbitrarily large. The principle
therefore introduces a new degree of relativity into fundamental 
physics beyond the familiar relativity of measuring sticks and time
pieces. It requires physics to be non-local on arbitrarily large 
lengthscales, yet conventional locality and causality must be recovered 
in everyday processes at low energy.

A detailed physical theory of black hole evaporation that incorporates
black hole complementarity has not been developed.\footnote{After this
lecture was delivered, Horowitz and Maldacena have proposed to reconcile
Hawking's semiclassical arguments with a unitary S-matrix by imposing a 
final state boundary condition at the black hole 
singularity \cite{Horowitz:2003he}. 
Their proposal may provide a realization of black hole complimentarity.}
There are indications, however, from string theory that such a 
description should be possible, especially in the context of the adS/cft 
correspondence.\footnote{We briefly describe the holographic
nature of this correspondence in Section~\ref{adscft}.} In this case,
the dual gauge theory is unitary and its S-matrix in principle 
includes processes where black holes are formed in the adS background
and subsequently evaporate \cite{Lowe:1999pk}. The problem is that the 
spacetime interpretation of gauge theory observables is obscure and it 
is non-trivial to establish local causality at low energy in adS 
spacetime in the language of the gauge theory, even in the absence of 
black holes \cite{Susskind:1998vk}. 

\section{The holographic principle}

Local quantum field theory leads to unitarity violation in the context of 
black hole evolution. This can be avoided by adopting the viewpoint of 
black hole complementarity but then spacetime physics is required to 
be non-local at a fundamental level. This notion of non-locality was
taken further by 't~Hooft \cite{'tHooft:1993gx} and Susskind 
\cite{Susskind:1995vu},
who argued that the number of available quantum states in any given
region is much smaller than one might naively expect. The claim is that
this number is not an extensive quantity, {\it i.e.} one that scales as the 
volume of the region in question, as one would find in any local quantum 
field theory with an ultraviolet cutoff, but is instead proportional to a 
surface area associated with the region. This dramatic reduction in 
the number of states as compared to conventional quantum field theory
is referred to as the holographic principle.  

\begin{figure}
\begin{center}
\psfrag{gamma}{$\Gamma$}
\psfrag{a}{$a$}
\includegraphics[width=4cm]{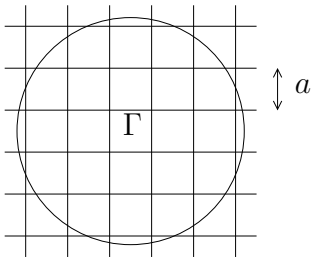}
\end{center}
\caption{\footnotesize Spherical region $\Gamma$ inside a cubic lattice.}
\label{latticespins}
\end{figure}

This principle only arises when gravitational effects are taken into
account. Let us illustrate this by a simple example.
Neglect gravity for the moment and consider a three-dimensional cubic
lattice of lattice spacing $a$ and with a spin located at each lattice site. 
Let $\Gamma$ be a spherical region as indicated in Figure~\ref{latticespins}. 
Let $V$ and $A$ be the volume and area of $\Gamma$ and assume that $V$ is 
large compared to $a^3$. Then the total number of spins inside $\Gamma$ is 
given by $V/a^3$ and if each spin is a two-level system the total number 
of states available to the spins in $\Gamma$ is $2^{V/a^3}$. The maximal 
entropy of the spin system in $\Gamma$ is given by the logarithm of the 
number of available states,
\begin{equation}
S_\textrm{\footnotesize max} = \frac{\log 2}{a^3}\,V \,.
\end{equation}
As expected, the maximal entropy is proportional to $V$. In fact, any 
local quantum field theory, regularized by an ultraviolet cutoff such as 
the lattice in our example, gives rise to a maximal entropy that is 
proportional to the spatial volume of the system.

This result turns out to be quite wrong when gravity is included.  To see
this, we carry out a thought experiment sometimes referred to as the
Susskind process. The system $\Gamma$ is placed at the center
of a large imploding shell of matter that carries energy just such that a
black hole of area $A$ is formed when the shell collapses into itself and
engulfs the spins in $\Gamma$.  We now compare the maximal entropy of the 
system, that consists of $\Gamma$ along with the collapsing shell, to that 
of the resulting black hole,

\begin{eqnarray}
S_{\textrm{\footnotesize initial}} &=& 
S_\Gamma + S_{\textrm{\footnotesize shell}}\,, \\
S_{\textrm{\footnotesize final}} &=& S_{\textrm{\footnotesize BH}} = 
\frac{1}{4}\, A \,.
\end{eqnarray}
By Bekenstein's generalized second law of thermodynamics the black hole
entropy must be the greater of the two and as a result the maximal 
entropy of the spin system in $\Gamma$ is bounded by $1/4$ of its 
area in Planck units,
\begin{equation}
S_{\textrm{\footnotesize final}} \ge S_{\textrm{\footnotesize initial}} 
\> \Rightarrow \> S_\Gamma \le  \frac{1}{4}\, A \,,
\end{equation}
instead of being proportional to the volume $V$.
Why does quantum field theory so grossly overestimate the maximal entropy?
The answer has to do with gravitational back reaction. Most of the states
of the spin system are highly excited and will curve the surrounding 
spacetime to the extent that it collapses to a large black hole, in fact 
much larger than the one formed in the Susskind process.

\begin{figure}
  \centering
  \psfrag{gamma}{$\Gamma$}
  \includegraphics[width=5cm]{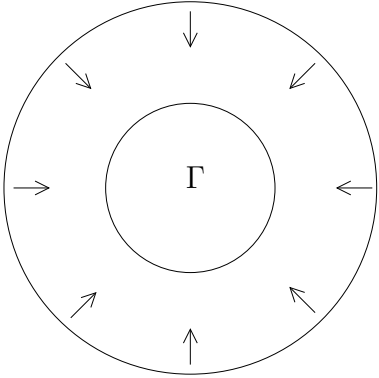}
  \caption{\footnotesize The Susskind process: The region $\Gamma$ is 
  placed inside an 
  imploding shell that forms a black hole with area $A$.}
  \label{implodingshell}
\end{figure}

\subsection{The spacelike entropy bound}
\label{spacebound}

A useful way to state the holographic nature of spacetime
physics is through entropy bounds that relate the number of available
states to a surface area associated with a given quantum system rather
than its volume.\footnote{For a detailed review of the holographic principle
and various entropy bounds see \cite{Bousso:2002ju}.}
We start with a simple entropy bound motivated by the analysis of the 
Susskind process. 

\begin{itemize}
\item
Spacelike entropy bound: {\it The entropy contained in any spatial region
will not exceed $1/4$ of the area of the region's boundary in Planck
units.}
\end{itemize}

This bound turns out to be too naive and is only valid under very 
restrictive assumptions. It is nevertheless useful as a step towards the 
much more universal covariant entropy bound of Bousso 
\cite{Bousso:1999cb} 
which we state later on.  It is instructive to consider some of the 
objections to the spacelike entropy bound. 

(i) {\it Particle species:} The entropy of a matter system depends on
the number of species of particles in the theory, and the entropy bound
will be violated if this number is sufficiently large. Just how large
the number has to be depends on the size of the system.  A simple 
estimate \cite{Bousso:2002ju} shows that the spacelike entropy bound fails 
for a weakly coupled gas in a volume with surface area $A$ in Planck 
units if the number of massless particle species is $N>A$.  
For a surface area of 1~cm$^2$ the required number of species is 
enourmous, $N\sim 10^{66}$. 
As there is no upper bound on the possible number of species in a field 
theory the entropy bound can formally always be violated. On the 
other hand, it is too much to expect an arbitrary field theory coupled to 
gravity to give sensible results. The holographic principle is put forward 
as a law of nature and the number of light matter fields in the real world 
is relatively small. 

Note also that we arrived at the spacelike entropy bound by analysing the 
Susskind process, which involves the formation of a black hole, but this 
black hole is violently unstable if $N$ exceeds the black hole area 
in Planck units, due to the large number of available channels for Hawking 
radiation~\cite{Wald:1999vt}.  We would therefore not expect the spacelike
entropy bound to hold in such a theory.

\begin{figure}
  \centering
  \psfrag{gamma}{$\Gamma$}
  \psfrag{B}{$B$}
  \includegraphics[width=4cm]{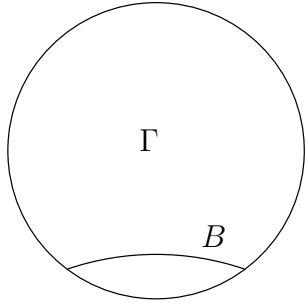}
  \caption{\footnotesize The region $\Gamma$ almost fills the closed 
  universe.}
  \label{southpole}
\end{figure}

(ii) {\it Closed FRW universe:} Consider a Friedmann-Robertson-Walker 
cosmological solution with positive spatial curvature. Spatial slices at
equal cosmic time have the geometry of a three-sphere. Take $\Gamma$ to
be a large comoving region centered around an observer at the 'north pole' 
of the three-sphere, as indicated in Figure~\ref{southpole}. Now let 
$\Gamma$ be so large that its boundary $B(\Gamma)$ approaches the south pole.
In this limit the region $\Gamma$ consists of the entire spatial universe 
while the area of its boundary shrinks to zero and the spacelike entropy 
bound is clearly violated.

(iii) {\it Spatially flat FRW universe:} 
Finally we consider another cosmological setup, where $\Gamma$ is taken
to be a spherical region of proper radius $R$ in a spatially flat FRW 
spacetime. The proper volume and surface area of this region are given by
\begin{equation}
V(\Gamma) = \frac{4\pi}{3}\, R^3 \qquad A(\Gamma) = 4\pi R^2 \,.
\end{equation}
Let us assume that this universe is filled with an isotropic and homogeneous 
background radiation carrying a uniform entropy density $\sigma$. Then the
total matter entropy contained in $\Gamma$ is 
\begin{equation}
S_\textrm{\footnotesize{matter}}(\Gamma) = \sigma V(\Gamma) = 
\frac{4\pi}{3}\, \sigma R^3 \,.
\end{equation}
For any non-vanishing entropy density the spacelike entropy bound is violated
when $R$ is sufficiently large,
\begin{equation}
R > \frac{3}{4\sigma} \> \Rightarrow \>
S_{\textrm{\footnotesize matter}}(\Gamma) > \frac{1}{4}A(\Gamma) \,.
\end{equation}

\subsection{Lightsheets and the covariant entropy bound}

The foregoing examples demonstrate the failure of the spacelike entropy 
bound but this does not mean that the holographic principle fails. 
What is needed is a more geometric entropy bound that adapts to dynamical 
situations like the ones in examples (ii) and (iii).  This is provided by 
Bousso's covariant entropy bound \cite{Bousso:1999cb}, which
involves light-cones rather than spacelike volume, but before we get to 
that we need to introduce a few geometric concepts.  We do this in the 
context of a simple example to avoid making the discussion too technical.

Consider a spherical surface $B$, with area $A(B)$, at rest in flat 
spacetime and imagine emitting light simultaneously from the entire 
surface at some time $t_0$. The light front will propagate in two directions, 
radially inwards and radially outwards.  One can also consider light
arriving radially at $B$ at time $t_0$ from the outside and inside 
respectively.  Thus there are four families of lightrays 
orthogonal to $B$,
\begin{center}
\begin{tabular}{cl}
+  + & future directed, outgoing, \\
+  $-$ & future directed, ingoing, \\
$-$  + & past directed, outgoing, \\
$-$  $-$ & past directed, ingoing.
\end{tabular}
\end{center}
There is nothing special about our spherical surface in this respect.
Every surface in Lorentzian geometry has four orthogonal light-like 
directions, two future directed and two past directed, and the 
corresponding families of lightrays trace out null hypersurfaces (at 
least locally) in spacetime. 

\begin{figure}
  \centering
  \psfrag{mm}{$--$}
  \psfrag{mp}{$-+$}
  \psfrag{pp}{$++$}
  \psfrag{pm}{$+-$}
  \psfrag{B}{$B$}
  \psfrag{time}{time}
  \includegraphics[width=7cm]{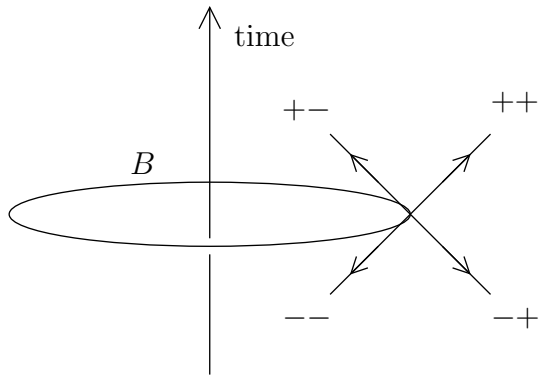}
  \caption{\footnotesize The circle represents a spherical surface 
           $B$ (with the polar angle suppressed). The lines denote future
           and past directed lightrays orthogonal to $B$.}
  \label{lightrays}
\end{figure}

The {\it expansion\/} $\theta$ of a family of lightrays, that are orthogonal 
to a surface, is positive (negative) if the rays are diverging (converging) 
as one moves along them away from the surface. In our simple example the sign 
of the expansion is easily determined.  The location at infinitesimal time 
$t=\Delta$ (or $t=-\Delta$ for past directed lightrays) as measured 
in the rest frame of $B$, of photons that were at $B$ at time $t=0$ 
defines a new surface $B'$ and we are interested in the area of $B'$ 
relative to $B$.  If $A(B')>A(B)$ the expansion of this particular family 
of lightrays is positive.
Beyond our example, the expansion of a family of lightrays orthogonal to 
any smooth surface in curved spacetime can be defined locally in a 
coordinate invariant manner \cite{Wald:1984rg} and its sign determined by 
comparing areas of neighboring surfaces intersected orthogonally by the 
lightrays.

The surface $B$ in our example is in a {\em normal\/} region where outgoing
lightrays (both future and past directed) have positive expansion and 
ingoing lightrays have negative expansion,
\begin{equation}
\theta_{++} > 0, \quad \theta_{+-} < 0, \quad \theta_{-+} > 0, 
\quad \theta_{--} < 0 \,.
\end{equation}
In a {\it future trapped} region on the other hand both the future
directed families of lightrays have negative expansion,
\begin{equation}
\theta_{++} < 0, \quad \theta_{+-} < 0, \quad \theta_{-+} > 0, 
\quad \theta_{--} > 0 \,.
\end{equation}
Such behavior is for example found in the collapsing region inside a black 
hole. There are other possibilities besides normal and future trapped but
one finds in all cases that at least two out of the four families of orthogonal 
lightrays have non-positive expansion, $\theta\leq 0$, locally
at the surface.  In degenerate cases this can be true of three or even all
four families.

\begin{figure}
  \centering
  \psfrag{B}{$B$}
  \psfrag{L}{$L$}
  \psfrag{Lp}{$L'$}
  \psfrag{time}{time}
  \includegraphics[width=5cm]{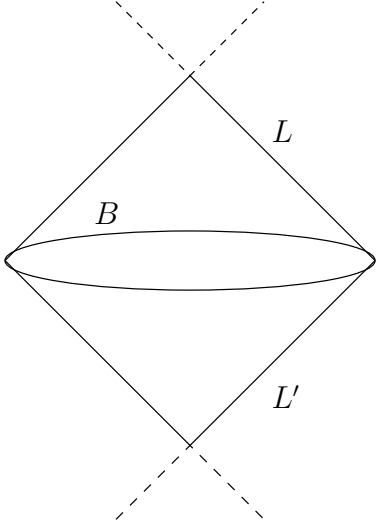}
  \caption{\footnotesize A lightsheet $L$ of a surface $B$ is a lightlike 
  hypersurface traced out by following a family of converging lightrays 
  orthogonal to $B$.}
  \label{hypersurface}
\end{figure}

A {\it lightsheet} $L$ of a surface $B$ is a lightlike hypersurface 
obtained by following a family of lightrays that is orthogonal to $B$ and 
has $\theta\leq 0$. It follows from our previous comment that every surface 
has at least two lightsheets.  The expansion will in general change as we
move along $L$ and the lightsheet ends where $\theta$ becomes positive.
For example, when converging lightrays self-intersect their expansion
turns positive and they no longer form a lightsheet. In other words, 
lightsheets do not extend beyond focal points as indicated in 
Figure~\ref{hypersurface} for our spherically symmetric example.

\begin{itemize}
\item
Covariant entropy bound: {\it The entropy on any lightsheet of a surface
$B$ will not exceed $1/4$ of the area of $B$ in Planck units}.
\end{itemize}

The covariant entropy bound appears to be universally applicable. At least 
there are no physically relevant counterexamples known.  
It has even been proven in the context of general relativity with the added
assumption that entropy can be described by a continuum fluid and with some
plausible conditions relating entropy density and energy 
density \cite{Flanagan:1999jp}.  Of course, as was pointed out in 
\cite{Flanagan:1999jp}, entropy at a fundamental level is not a fluid and 
the assumed conditions relating entropy and energy are not aways satisfied.  
It should therefore be stressed that the covariant
entropy bound has not been derived from first principles, after all the first 
principles of quantum gravity are unknown, but is rather an observation about 
the nature of matter and gravity that should be explained by a fundamental 
theory.

Under certain assumptions the covariant entropy bound implies the spacelike
bound of Section~\ref{spacebound} \cite{Bousso:1999cb} but the covariant bound
is valid much more generally. It is instructive to see how the covariant
bound deals with the various objections that were offered to the spacelike 
bound.

(i) {\it Particle species}: The species problem is the same as in the
spacelike case and is equally relevant (or irrelevant) here.

(ii) {\it Closed FRW universe}: Lightrays directed towards the north pole
of the three sphere from $B$ in Figure~\ref{southpole}, {\it i.e.} ones
that traverse $\Gamma$, have positive expansion. Therefore they do
not trace out a lightsheet and the covariant entropy bound does not
apply. Both lightsheets of $B$ are directed towards the south pole in
this case and the covariant entropy bound is valid for the complement of
$\Gamma$.

(iii) {\it Spatially flat FRW universe}:
Without going into details we note that the problem here had to do
with very large regions in a spatially flat universe. The 
surface of such a region will have a past directed lightsheet but
this lightsheet terminates at the initial singularity and has therefore
much less entropy on it than is contained in the original spatial volume.
It can be shown that the entropy that passes through the lightsheet
is less that a quarter of the surface area of the region under
consideration in Planck units \cite{Fischler:1998st}.

\section{The adS/cft correspondence} 
\label{adscft}

The holographic principle is put forward as a basic principle of 
physics and as such it should be manifest in any successful fundamental 
theory. It is therefore natural to ask to what extent superstring theory,
the leading candidate for a unified quantum theory of matter and 
gravity, is holographic. Since string theory is far from a finished 
product, with major conceptual problems unsolved, it may be premature
to subject it to this test.  Yet, remarkably, it has already produced
a setting where holography is explicitly realized. We finish this 
lecture with a quick sketch of the adS/cft correspondence and an 
order-of-magnitude estimate showing how the number of degrees of
freedom in this nonperturbative definition of string theory is in
line with the holographic principle.

The setting for the original adS/cft correspondence \cite{Maldacena:1998re} 
is the physics of $N$ coincident Dirichlet three-branes in ten-dimensional 
spacetime.  This physics is governed by an action of the form 
\begin{equation}
S = S_\textrm{\footnotesize bulk}  + S_\textrm{\footnotesize brane}
 + S_\textrm{\footnotesize int}\,.
\end{equation}
Here $S_\textrm{\footnotesize bulk}$ is the ten-dimensional gravitational 
action, {\it i.e.} type IIB supergravity along with $\alpha'$ corrections 
from string theory, while $S_\textrm{\footnotesize brane}$ is the worldvolume
action of the $N$ Dirichlet-branes, {\it i.e.} $d=4$, $\mathcal{N}=4$ 
supersymmetric SU$(N)$ Yang-Mills theory along with its own $\alpha'$
corrections.  $S_\textrm{\footnotesize int}$ describes the coupling
between the ten-dimensional bulk and the branes and this coupling can be 
ignored in the limit of weak string coupling, $N g_s \ll 1$. 

A stack of coincident Dirichlet three-branes has a dual description as an 
extended object in supergravity. The corresponding ten-dimensional metric 
is given by the line element
\begin{equation}
ds^2 = \frac{1}{\sqrt{H(r)}} \left(-dt^2 + \sum_{i=1}^{3} dx_i dx_i \right)
      + \sqrt{H(r)}\left( dr^2 + r^2 d\Omega_5^2 \right) \,,
\end{equation}
where $H(r) = 1+ \frac{R^4}{r^4}$. The parameter $R$ is a characteristic 
length which must be large compared to the string length $\sqrt{\alpha'}$ 
in order for the supergravity solution to be valid. The parameters of the 
two descriptions are related through
\begin{equation}
\label{rnref}
R^4 = 4\pi g_s \alpha '^2 N \,,
\end{equation}
so the supergravity description requires $g_s\ll 1\ll Ng_s$.

The three-brane geometry has an event horizon at $r=0$ in these 
coordinates, and due to gravitational redshift the near horizon region 
$r\ll R$ is effectively decoupled from the bulk ten-dimensional supergravity.
The metric of the near-horizon region is given by 
\begin{equation}
ds^2 \simeq R^2\left[
   z^2\left(-d\tilde{t}^2 + \sum_{i=1}^{3} d\tilde{x}_i d\tilde{x}_i \right)
      + \frac{dz^2}{z^2} +d\Omega_5^2
         \right] \,,
\end{equation}
where we have introduced dimensionless variables through $r=Rz$, 
$t=R\tilde{t}$ and $x_i=R\tilde{x}_i$. This is the metric of 
adS$_5\times$S$_5$ using Poincar\'e coordinates for the adS$_5$ part.

The adS/cft correspondence follows from the observation that the 
above two descriptions of the D3-brane system both involve two decoupled 
factors, and in each case one of the factors is bulk ten-dimensional 
supergravity. By identifying the other factors with each other we
are led to a duality between the worldvolume SU$(N)$ gauge theory, which
is a four-dimensional conformal field theory, and string theory in the 
adS$_5\times$S$_5$ near-horizon geometry. The two dual descriptions apply 
at different coupling strength, $N g_s \ll 1$ {\it vs.} $g_s\ll 1\ll Ng_s$. 

\subsection{AdS/cft and the holographic principle}
\label{adsinfo}

We now give an argument, due to Susskind and Witten \cite{Susskind:1998dq}, 
that the number of degrees of freedom in adS-gravity in fact satisfies a 
holographic bound. For this it is more convenient to adopt so-called cavity
coordinates for adS$_5$,
\begin{equation}
\label{cavity}
ds^2 = R^2\left[
-\left(\frac{1+u^2}{1-u^2}\right)d\tau^2
+\frac{4}{(1-u^2)^2}\left( du^2 + u^2d\Omega_3^2\right)+d\Omega_5^2
\right] \,.
\end{equation}
The total entropy of the gravitational system is infinite because the 
spatial proper volume of adS spacetime diverges. 
In cavity coordinates the spatial boundary of the adS geometry is located 
at $u=1$ and we impose a cutoff at $u = 1- \epsilon$ with $\epsilon \ll 1$.
With this infrared regulator in place the spatial volume is finite.
The proper area of the spatial boundary is also finite and can easily be
obtained from (\ref{cavity}). This leads to the following holographic
entropy bound,
\begin{equation}
\label{holobound}
S \le \frac{1}{4} \times \textrm{``area''}
  \sim \left( \frac{R}{\epsilon}\right)^3 R^5
    = \frac{R^8}{\epsilon^3} \,,
\end{equation}
up to factors of order one.

The dual field theory is scale invariant and so it also has infinite entropy.
This time around the divergence is an ultraviolet effect and can be regulated
by introducing a short-distance cutoff. An important feature of the adS/cft 
correspondence is that infrared effects in adS space appear in the
ultraviolet in the dual gauge theory. We therefore take the short
distance cutoff in the field theory to be proportional to $\epsilon$. 
SU$(N)$ Yang-Mills theory has $O(N^2)$ ``gluon'' fields so the total entropy
in the dual gauge theory then goes like
\begin{equation}
S \sim \frac{N^2}{\epsilon^3} \,,
\end{equation}
which is seen to saturate the holographic bound (\ref{holobound})
when we take into account the relation (\ref{rnref}) between $R$ and $N$. 

\section{Discussion}

We reviewed the black hole information problem and argued that
developments in string theory strongly favor its resolution in terms
of unitary evolution. This comes at a price of introducing a
fundamental nonlocality into physics, but historically this nonlocality 
in black hole evolution served to motivate the holographic principle, 
a far reaching new paradigm for quantum gravity.

Many intresting aspects of holography were not touched upon here.
The goal was to convey some of the basic ideas rather than give a 
survey of the field, which is by now quite wide. Recent work has included
a revival of interest in matrix models of two-dimensional gravity
\cite{reloaded}, which provide a relatively simple realization of the 
holographic principle, and also the application of holographic ideas to 
cosmology \cite{susban}. 

\vspace{10pt}
This work was supported in part by grants from the Research Fund of the 
Icelandic Science and Technology Policy Council and the University of 
Iceland Research Fund.

\providecommand{\href}[2]{#2}\begingroup\raggedright\endgroup

\end{document}